\documentclass[aps,prl,twocolumn]{revtex4}
\usepackage{graphicx}% Include figure files
\usepackage{bm}% bold math
\begin{document}
\def \tr{{\mbox{tr~}}}
\def \ra{{\rightarrow}}
\def \ua{{\uparrow}}
\def \da{{\downarrow}}
\def \be{\begin{equation}}
\def \ee{\end{equation}}
\def \ba{\begin{array}}
\def \ea{\end{array}}
\def \bea{\begin{eqnarray}}
\def \eea{\end{eqnarray}}
\def \nn{\nonumber}
\def \l{\left}
\def \r{\right}
\def \half{{1\over 2}}
\def \etal{{\it {et al}}}
\def \cH{{\cal{H}}}
\def \cM{{\cal{M}}}
\def \cN{{\cal{N}}}
\def \cQ{{\cal Q}}
\def \cI{{\cal I}}
\def \cV{{\cal V}}
\def \cG{{\cal G}}
\def \bS{{\bf S}}
\def \bI{{\bf I}}
\def \bL{{\bf L}}
\def \bG{{\bf G}}
\def \bQ{{\bf Q}}
\def \bR{{\bf R}}
\def \br{{\bf r}}
\def \bu{{\bf u}}
\def \bq{{\bf q}}
\def \bk{{\bf k}}
\def \bz{{\bf z}}
\def \bx{{\bf x}}
\def \tJ{{\tilde{J}}}
\def \W{{\Omega}}
\def \e{{\epsilon}}
\def \lam{{\lambda}}
\def \L{{\Lambda}}
\def \a{{\alpha}}
\def \t{{\theta}}
\def \b{{\beta}}
\def \g{{\gamma}}
\def \D{{\Delta}}
\def \d{{\delta}}
\def \w{{\omega}}
\def \s{{\sigma}}
\def \f{{\varphi}}
\def \x{{\chi}}
\def \e{{\epsilon}}
\def \h{{\eta}}
\def \G{{\Gamma}}
\def \z{{\zeta}}
\def \hatt{{\hat{\t}}}
\def \hn{{\bar{n}}}
\def \vk{{\bf{k}}}
\def \vq{{\bf{q}}}
\def \gk{{\g_{\vk}}}
\def \nd{{^{\vphantom{\dagger}}}}
\def \yd{^\dagger}
\def \av#1{{\langle#1\rangle}}
\def \ket#1{{\,|\,#1\,\rangle\,}}
\def \bra#1{{\,\langle\,#1\,|\,}}
\def \braket#1#2{{\,\langle\,#1\,|\,#2\,\rangle\,}}

\title{Superfluid-insulator transition in a moving system of interacting bosons}
\author{E.~Altman, A.~Polkovnikov, E.~Demler, B. I.~Halperin, and M.~D.~Lukin}

\address{Physics Department, Harvard University, Cambridge, MA 02138}
\date{\today}
\begin{abstract}
We analyze stability of superfluid currents in a system of
strongly interacting ultra-cold atoms in an optical lattice. We
show that such a system undergoes a dynamic, irreversible phase
transition at a critical phase gradient that depends on the
interaction strength between atoms. At commensurate filling, the
phase  boundary continuously interpolates between the classical
modulation instability of a weakly interacting condensate and the
equilibrium quantum phase transition into a Mott  insulator state
at which the critical current vanishes. We argue that quantum
fluctuations smear the transition boundary in low dimensional
systems. Finally we discuss the implications to realistic
experiments.
\end{abstract}

\maketitle

The quantum phase transition from superfluid (SF) to Mott
insulator (IN) ~\cite{fisher} is an important paradigm of strongly
correlation physics. Recently this transition was demonstrated in
spectacular experiments involving ultra cold atoms in an optical
lattice ~\cite{zoller,greiner}. An important feature of these
systems is that they can be essentially isolated from the
environment, opening unique possibilities to study, not only the
equilibrium phase diagram, but also quantum dynamics very far from
equilibrium
~\cite{mandel,inguscio,niu,smerzi,ehud,anatoli,charles}. In
particular, it is now possible to explore a new class of phenomena
involving non-equilibrium dynamics in the vicinity of quantum
phase transitions.

In this letter we analyze the stability of superfluid current flow
in a system of strongly interacting bosons in a lattice. We show that at the
mean field level such a system, undergoes a dynamic phase
transition, associated with irreversible decay of the superfluid flow at a
critical momentum that depends on the interaction strength.
%Quantum fluctuations allow decay of current below the
%critical momentum via instanton (phase slip) tunneling\cite{coleman}.
We argue that quantum phase fluctuations play an important role
near the phase boundary. In systems of lower dimensionality, they
broaden the transition significantly as was indeed observed in recent
experiments~\cite{chad} and numerical
simulations~\cite{charles,nist} in one dimensional systems. In
three dimensions, by contrast, we predict that the current decay
rate exhibits a sharp discontinuity at the mean field transition.
We propose to test this prediction with transport experiments in
three dimensional optical lattices ~[\onlinecite{greiner}].

A well studied effect, closely related to our discussion, is the
modulational instability of weakly interacting bosons on a lattice
\cite{niu,smerzi}.  It was experimentally demonstrated
~\cite{inguscio, oberthaler} that such a condensate undergoes a
dynamical localization transition, involving onset of chaos, when
the phase gradient (or condensate momentum) associated with the
flow exceeds $\pi/2$ per lattice unit. The dynamic phase
transition described in this letter interpolates continuously
between the classical instability at condensate momentum $\pi/2$
and the quantum phase transition into the Mott state at zero
current, thereby establishing a natural connection between the two
transitions.

\begin{figure}[ht]
% \centering
\includegraphics[width=8cm]{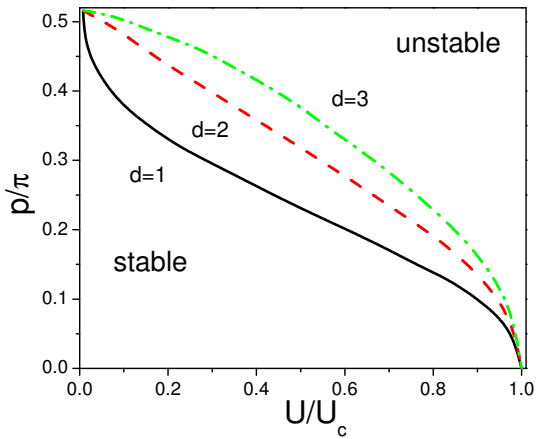}
\caption{Stability phase diagram in the plane of phase twist per
bond  versus dimensionless interaction for filling of $N=1$
particle per site found from numerical solution of time dependent
Gutzwiller equations (\protect{\ref{TDG}}).}
 \label{fig:instab}
\end{figure}

Dependence of the critical momentum on the interaction can be
understood as follows. The superfluid current $I$ associated with
a condensate moving within the lowest Bloch band is $I(p) \sim
\rho_s\sin(p)$, where $p$ is the (quasi)momentum of the condensate
measured in the units of inverse lattice constant and $\rho_s$ is
the superfluid density. In a weakly interacting condensate
$\rho_s$ is independent of $p$. Thus, the current increases with
$p$ up to a maximal value at $p_c = \pi/2$. Beyond this point, the
effective mass changes sign and any further increase in $p$
results in decrease of the current, rendering the superfluid
unstable~\cite{decay}. At strong interactions $\rho$ itself is a
function of the effective mass and thus also of $p$. Specifically
$\rho$ decreases as $p$ is increased, such that the maximum of
$I(p)$ occurs in general at $p_c<\pi/2$. In particular, close to
the Mott insulator, a slight increase in effective mass, and thus
also in $p$ leads to vanishing of $\rho$. Therefore $p_c$ tends to
zero toward the Mott transition. These considerations include the
effect of quantum depletion at the mean field level, which is
effective in all dimensions, and lead to a stability phase diagram
(for commensurate filling) depicted in  Fig. \ref{fig:instab}.
Smearing of the transition lines by quantum phase slips beyond
mean field theory, is effective only in lower dimensionality.

%%%%%%%%%%%%%%%%%%%%%%%%%%%%%%%%%%

Static and dynamic properties of condensates in optical
lattices are described by the Bose Hubbard model (BHM)
\be
H=\frac{U}{2}\sum_i n_i(n_i-1)-J\sum_{\langle ij\rangle} (a\yd_i
a\nd_j+\mbox{H.c.}),
\label{BHM}
\ee
where $J$ is the hopping amplitude between the nearest neighbors
$\langle ij\rangle$, $U$ is the on-site repulsive interaction, and
$n_i=a_i^\dagger a_i$ is the number operator. We denote the
average number of bosons per site by $N$. Provided  the system is
deep in the superfluid phase ($JN\gg U$), the dynamics can be well
approximated by the discrete GP equations
\be
i{d\psi_i\over d t}=-J\sum_{k\in O}\psi_{k}+U|\psi_i|^2\psi_i,
\label{gp}
\ee
where $\psi_i=\av{a_i}$ is the matter field and the set $O$
contains the nearest neighbors of site $i$. Linear mode analysis
around the stationary current carrying solutions $\psi_i=
\sqrt{N}\exp(ipx_i)$, yields the onset of instability at
$p=\pi/2$~\cite{niu,smerzi}.

Close to the Mott transition, increased quantum fluctuations
invalidate the GP description. However, one can still use
semiclassical order parameter dynamics if one coarse grains the
system into blocks of roughly a coherence length $\xi$, which is
related to the superfluid density by a standard scaling
form~\cite{sachdev}. As in the weak coupling theory, the current
should become unstable when the phase change per unit cell exceeds
$\pi/2$, with the unit cell now of order $\xi$. Since the
coherence length, $\xi$, diverges at the transition, we expect
$p_c$ to vanish as $1/\xi$, as we indeed find below. The diverging
length scale facilitates a continuum description of the dynamics
close to the Mott transition, in the form of a time dependent
Ginzburg-Landau equation~\cite{ehud,decay}
\be
\ddot\psi=\nabla^2\psi+\psi(\xi^{-2}-|\psi|^2).
\label{tdgl}
\ee
Within the mean-field approximation, and  in the limit of large
average occupation $N\gg 1$,  $\xi^{-2}=2d(1-u)$, where $u=U/(8JN
d)$ is the dimensionless interaction constant. If $N$ is not too
large then Eq.(\ref{tdgl}) still holds but the expressions for
$\xi$ and $u$ are more complicated~\cite{decay}.

We choose the zero of energy at the state with integer filling.
Then the deviation from commensurate density is given in terms of
the superfluid order parameter $\psi$ by $\d
n=C_d(\psi^\star\dot\psi-\psi\dot\psi^\star)/2i$, with
$C_d=u^{-1}(2d)^{-3/2}$ and $d$ being the dimensionality of the system.
Note that $\int d^d x\, \d n$ is a constant of motion under
(\ref{tdgl}). The supercurrent is similarly given by
$I=C_d(\psi^\star\nabla\psi-\psi\nabla\psi^\star)/2i$.

The equation of motion (\ref{tdgl}) admits uniform solutions
$\psi=\rho e^{i p x+i\mu t}$, with $\rho=\sqrt{\xi^{-2}+\mu^2-p^2}$,
which are characterized by a phase gradient $p$ and a relative
density
\be
\d n=C_d\mu(\xi^{-2}+\mu^2-p^2).
\label{deltan}
\ee
In particular, at commensurate filling $\mu=0$ and $\psi$ is
time-independent. To analyze whether these solutions are stable we
find the spectrum of small fluctuations around them. There are two
eigenmodes in the superfluid regime ($\xi^{-2}>0$): a stable
gapped mode, and a phase (Bogoliubov) mode with linear
dispersion at long wave lengths. The spectrum of the latter, for
wavevectors parallel to the current, reads:
\be
\w(k)={2\mu p\over 2\mu^2+\rho^2} k+{\rho \over
2\mu^2+\rho^2}\sqrt{\xi^{-2}+3\mu^2-3p^2}|k|.
\label{wk}
\ee
Here the first term is analogous to the  usual Doppler shift and
the second describes propagation of the sound waves in the moving
reference frame. The onset of imaginary frequencies marking the
instability occurs at $p_c^2-\mu(p_c)^2=1/3\xi^2$. Combining this
with Eq. (\ref{deltan}) we find that for $N\gg 1$
\be
p_c=\sqrt{\frac{1}{3\xi^2}+\left(\frac{3\d n \xi^2}{4 C_d}\right)^2}
\label{pc}
\ee
%\be
%p_c=\sqrt{\frac{d}{2}\left(\frac{3u\d n}{1-u}\right)^2+{2\over 3}d(1-u)}.
%\label{pc}
%\ee
As argued before on general grounds, in the case of commensurate filling
$\d n=0$, the critical phase
gradient vanishes toward the equilibrium Mott transition ($u=1$)
as $p_c\propto 1/\xi\propto \sqrt{1-u}$.
At incommensurate density there is no equilibrium Mott transition.
As a result, we do not expect the instability to reach $p=0$.
Indeed, $p_c$ has a minimum at $u<1$ ($\xi<\infty$), and diverges as $u\to 1$.
The divergence, simply signals the breakdown of the continuum theory
and is cutoff by the lattice.

To interpolate between the regimes of weak and strong interactions
we employ the Gutzwiller approximation~\cite{kotliar}. In this
approach, the wavefunction is assumed to be factorizable:
\be
\ket{G}=\prod_j \left[\sum_{n=0}^\infty f_{jn}\ket{n}_j\right].
\label{G}
\ee
Here $j$ is a site index and $n$ is the site occupation. The
ansatz (\ref{G}) supplemented by self-consistency conditions leads
to equations of motion for the variational parameters:
\begin{eqnarray}
&&-i\dot{f}_{jn}=\frac{U}{2} n(n-1)f_{jn}-\nonumber\\
&&-Jz(\sqrt{n} f_{j, n-1}\psi_j +\sqrt{n+1}
f_{j,n+1}\psi^\star_j),
\label{TDG}
\end{eqnarray}
where
\be
\psi_j\equiv{1\over z}\sum_{i\in O}\bra{G}a_i\ket{G}.
\ee
For actual calculations we truncated (\ref{G})
at five and ten states per site without noticeable
differences in the results.

Equations (\ref{TDG}), admit uniform current carrying solutions.
We numerically check their stability to slight perturbations in
the equations of motion. We show the stability boundaries at
commensurate filling in Fig.~\ref{fig:instab}. It is evident that
the dynamical instability at $\pi/2$ in the GP regime is
continuously connected to the equilibrium (zero current) Mott
transition.

\begin{figure}[h]
% \centering
\includegraphics[width=8cm]{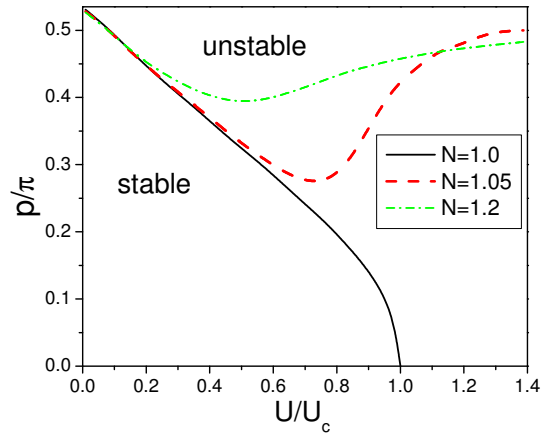}
\caption{Stability phase diagrams for different filling factors in
a two-dimensional lattice from numerical solution of the time
dependent Gutzwiller equations (\protect{\ref{TDG}}). Away from
commensurate filling the critical phase gradient reaches a minimum
and climbs back to $\pi/2$.}
\label{fig:instab1}
\end{figure}

We  perform a similar analysis at incommensurate filling
(Fig.~\ref{fig:instab1}). In agrement with the continuum
expression (\ref{pc}) we find that the critical momentum $p_c$
reaches a minimum at some $u<1$. At stronger interactions, $p_c$
increases, and saturates at $\pi/2$.

%%%%%%%%%%%%%%%%%%%%%%%%%%%%%%%%%%

The mean field transition discussed above ignores the possibility
of current decay below the critical momentum due to quantum
tunneling out of the metastable state. Such processes are
exponentially suppressed by a tunneling action through a barrier.
But since the barrier vanishes at the mean field instability, they
can potentially broaden the transition rendering the phase
diagrams of Figs. \ref{fig:instab} and \ref{fig:instab1}
meaningless. This problem is addressed in full detail in Ref.
~\cite{decay} within the general framework of ~\cite{coleman}.
Here we one important result for the decay rate close to the
equilibrium SF-IN transition, i.e. at small critical current. To
obtain the tunnelling action close to the critical current we
expand the GL action associated with (\ref{tdgl}) around the
metastable solution to cubic order in the fluctuations. Then we
use a scaling approach similar to the one introduced in
Ref.~[\onlinecite{Kline}] in the context of spinodal
decomposition. Since the instability in (\ref{wk}) arises at
$\bk\to 0$ the barrier vanishes only for a tunneling instanton of
diverging (space-time) volume. By dimensional analysis  we find
$V_{inst}\propto (p_c-p)^{d+1/2}$, while the energy density of the
barrier $E_b\propto (p_c-p)^3$. Consequently the instanton action
is $S_{inst}\sim E_b\times V_{inst}\propto B_d (p_c-p)^{2.5-d}$.
Note that this scaling is different from that derived in
Ref.~[\onlinecite{Kline}], because the field $\psi$ is complex.
Thus in one and two dimensions the tunneling action vanishes
continuously toward the mean field transition. Then we expect the
transition to become a wide crossover ($B_d$ is calculated in
~\cite{decay} and found to be of order unity). This agrees with
recent experiments~\cite{chad} and numerical
simulations~\cite{charles} in one dimensional traps.

In three dimensions, by contrast, the action of ``critical''
instantons diverges because of their diverging volume. Current
would then rather decay via non critical instantons, i.e. ones of
finite size that feel a finite energy barrier, and thus cost
finite action. We therefore predict that in three dimensions, the
decay rate will exhibit a discontinuity at the mean field
transition. A variational calculation~\cite{decay} yields a rate
$\Gamma\propto e^{-4.3}$ in the vicinity of the transition. In
this sense the mean-field phase diagram (Fig.~\ref{fig:instab}) is
well defined in three dimensions.

In realistic experimental situations condensates are confined in
harmonic traps which  leads to a non uniform density distribution
in the form of domains with different $N$. In the weakly
interacting  regime the critical momentum is $p_c=\pi/2$, i.e. it
is  insensitive to the spatial density variation induced by the
harmonic confinement. By contrast, in the regime of strong
interactions the position of the dynamical instability strongly
depends on the filling factor $N$ that is directly affected by the
density distribution. In particular, the motion first becomes
unstable for the smallest integer filling $N =1$.

\begin{figure}[h]
% \centering
\includegraphics[width=8cm]{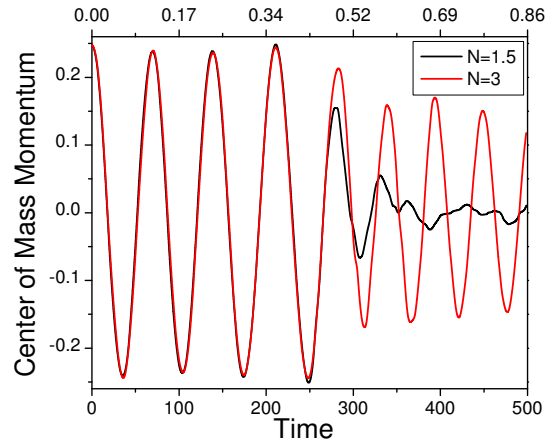}
\caption{Time dependence of the condensate momentum in a
two-dimensional harmonic trap with different filling factors per
central site. The simulated system is a lattice of dimensions $120\times 60$
with global trapping potential $V(j_x,j_y)=0.01(j_x^2+j_y^2)$.
We set the hopping amplitude $J=1$ while increasing the interaction
linearly in time: $U(t)=0.01 t$.}
\label{fig:trap}
\end{figure}

In Fig.~\ref{fig:trap} we plot the time evolution of the
condensate momentum (computed within the Gutzwiller approximation)
in a trap, for two different filling factors. The center of mass
motion becomes unstable at approximately the same interaction
strength in both cases. But while at smaller filling the
condensate motion rapidly becomes chaotic as in the uniform case,
damping of oscillations at larger filling occurs much more
gradually. These results can be understood by noting that if the
phase gradient in the condensate exceeds the critical value
corresponding to $N=1$ these domains become unstable triggering
the decay of current. However, when there is high filling of the
central sites the overall weight of domains with $N=1$, is small.
Thus the effect of the instability on the total current is
reduced.

%%%%%%%%%%%%%%%%%%%%%%%%%%%%%%%%%%

An important experimental manifestation of these results is the
inherently irreversible nature of the phase transition at finite
currents. Consider a situation in which a moving condensate is
first prepared on a weak lattice. Then, the depth of the periodic
potential is increased adiabatically~\cite{unote}, which
corresponds to moving along a horizontal line in the parameter
space of Fig.~(\ref{fig:instab}). Finally, the lattice depth is
slowly decreased back to its original state and the visibility of
the interference fringes compared to their initial value. If in
this sequence we pass the instability, then the current will decay
into incoherent excitations and heat the condensate. This will
result in total loss of current and reduced visibility of the
interference fringes at the end of the cycle. Such experiments
could be used to probe the nonequilibrium phase diagram
(Fig.~{\ref{fig:instab}}) and to determine the position of the
equilibrium Mott transition.

This work was supported by the NSF (grants DMR-01328074,
DMR-0233773, DMR-0231631,  DMR-0213805,  PHY-0134776), the Sloan
and the Packard Foundations, and by Harvard-MIT CUA.


\begin{references}
\bibitem{unote} One should be aware that
if in a harmonic trap the tunneling (not the interaction strength)
decrease in time then the amplitude of momentum oscillations
increases (see Ref.~[\onlinecite{decay}] for the details).
\end{references}
\end{document}